\documentclass[usenatbib]{basi}

\usepackage[T1]{fontenc}
\usepackage[british]{babel}
\usepackage[varg]{txfonts}
\usepackage{natbib}
\usepackage{deluxetable}
\makeatletter\NAT@longnamesfalse\makeatother 

\begin{document}

\title[Accretion onto White Dwarfs]{Theoretical Studies of Accretion of Matter onto White Dwarfs\\
 and the Single Degenerate Scenario for Supernovae of Type Ia}

\author[S. Starrfield, et al. ]{S. Starrfield$^1$\thanks{email: \texttt{starrfield@asu.edu}},
C. Iliadis$^2$\thanks{email:\texttt{iliadis@unc.edu}},
F. X. Timmes$^1$\thanks{email:\texttt{fxt44@mac.com}},
W. R. Hix$^3$\thanks{email:\texttt{raph@utk.edu}},
W. D. Arnett$^4$\thanks{email:\texttt{darnett@as.arizona.edu}},
C. Meakin$^5$\thanks{email:\texttt{casey.meakin@gmail.com}},
W. M. Sparks$^5$\thanks{email:\texttt{warrensparks@comcast.net}}\\
$^1$School of Earth and Space Exploration, P.O. Box 871404, Arizona State University, Tempe, AZ 85287-1404, USA\\
$^2$Dept. of Physics \& Astronomy,  University of North Carolina, Chapel Hill, NC 27599-3255\\
$^3$Dept. of Physics and Astronomy, University of Tennessee, Knoxville, TN 37996-1200\\
$^4$ Dept. of Astronomy, University of Arizona, Tucson, AZ, 85721\\
$^5$ Los Alamos National Laboratory, Los Alamos, NM, 87545}

\pubyear{2012}
\volume{00}
\pagerange{\pageref{firstpage}--\pageref{lastpage}}

\date{Received --- ; accepted ---}

\newcommand{\apj}{ApJ}
\newcommand{\nat}{Nature}
\newcommand{\aj}{AJ}
\newcommand{\gca}{Geochemica \& Cosmica Acta}
\newcommand{\apjl}{ApJL}
\newcommand{\pasp}{PASP}
\newcommand{\aap}{A\&A}
\newcommand{\mnras}{MNRAS}
\newcommand{\apjs}{ApJSupp}
\newcommand{\araa}{ARAA}
\newcommand{\aapr}{A\&A Reviews}
\maketitle

\label{firstpage}

\begin{abstract}
We review our current knowledge about the
thermonuclear processing that occurs during the evolution of 
accretion onto white dwarfs both with and without the mixing of core with accreted
material.  We present a brief summary of the Single Degenerate Scenario for the
progenitors of Type Ia Supernovae in which it is assumed that a low mass carbon-oxygen
white dwarf is growing in mass as a result of accretion from a secondary star in a 
close binary system.  The growth in mass requires that
more material remain on a white dwarf after a thermonuclear runaway than is ejected by the explosion.
Recent hydrodynamic
simulations of accretion of solar material onto white dwarfs {\it without mixing} always produce a thermonuclear runaway
and ``steady burning'' does not occur.  For a broad range in WD mass (0.4 M$_\odot$  to 1.35 M$_\odot$),
the maximum ejected material occurs 
for the 1.25M$_\odot$  sequences and then decreases as the white dwarf mass decreases.  
Therefore, the white dwarfs are growing in mass as a consequence of the accretion of solar material
and as long as there is no mixing of accreted material with core material.   In contrast, a 
thermonuclear runaway in the accreted hydrogen-rich
layers on the {\it low} luminosity WDs in close binary systems where mixing of core matter with accreted material has occurred
is the outburst mechanism for Classical,
Recurrent, and Symbiotic novae.   The differences in characteristics of
these systems is likely the WD mass and mass accretion rate.  
The high levels of enrichment of
CN ejecta in elements ranging from carbon to sulfur
confirm that there is dredge-up of matter from the core of the WD
and enable them to contribute to the chemical
enrichment of the interstellar medium.  Therefore, studies of
CNe can lead to an improved understanding of Galactic
nucleosynthesis, some sources of pre-solar grains, and the
Extragalactic distance scale.  The
characteristics of the outburst depend on the white dwarf mass,
luminosity, mass accretion rate, and the chemical composition of
both the accreting material and WD material.  The properties of the outburst also depends on
when, how, and if the accreted layers are mixed with the WD core and the mixing mechanism is still unknown. 

\end{abstract}


\section{Introduction}\label{sec:intro}

We take, as given,
the model for Cataclysmic Variables (CVs) that has been developed over the past 50 years \citep{kraft_1956_aa}:
accretion occurs onto the white dwarf  (WD) component in a
close binary system.  While the compact object is a WD, the secondary is 
a larger, cooler star that typically fills its Roche Lobe (in the Restricted Three-Body Problem).  
In some Symbiotic systems,
which include the Symbiotic Novae such as RS Oph, V407 Cyg, and T CrB, the orbital period
is so long that the secondary may not fill its Roche Lobe but mass transfer onto the WD must be occurring
in order for an explosion to occur.   
We note, in addition, that larger only refers to the radius of the secondary. 
In many cases, the secondary is less massive than the WD.   
Because the secondary fills its Roche Lobe,  any
tendency for it to grow in size because of evolutionary processes or for
the Roche Lobe to shrink because of angular momentum losses will cause a flow of
gas through the inner Lagrangian point into the Roche Lobe surrounding the WD.
The size of the WD is small compared to the size of its Lobe and
the angular momentum of the transferred material causes it to spiral
into an accretion disc surrounding the WD.  Some viscous process
 acts to transfer material inward and angular momentum outward so
that some fraction of the material lost by the secondary ultimately ends up on
the WD while some material must be ejected from the accretion disk.

The accreting material gradually forms a layer of fuel on the WD and
the bottom of this layer is compressed by the strong gravity of the WD and the continuously
infalling material.  The accreted layer is heated both by compression and by the flow of
heat from the interior.  Over a period of time, the accreted layer on the WD grows in
mass until the bottom reaches temperatures that are sufficiently high to
initiate thermonuclear burning of hydrogen by the proton-proton chain of
reactions.  In addition, by this time the bottom of the layer has become electron
degenerate.   Because the material is degenerate,  the energy release does not result
in expansion and cooling of the accreted layers. 
Therefore, once nuclear burning in the bottom of the
layer reaches thermonuclear runaway (TNR) conditions, the
temperatures in the nuclear burning region will exceed $10^8$K
for low mass WDs (M$_{\rm{WD}}$$ < $1.0 M$_{\odot}$).  For higher mass WDs (M$_{\rm{WD}}$$ > $1.2 M$_\odot$) these
temperatures can exceed $2 \times 10^8$K  even if there has been no mixing of core
material with the accreted material.  
Hydrodynamic studies show that the further evolution of the thermonuclear burning and its evolution
to a TNR on the WD
now depends upon the mass and luminosity of the WD, the rate
of mass accretion, and the chemical composition of the nuclear burning layer. 
This last condition implies that the evolution of the resulting TNR will differ depending on whether or not
mixing of core with accreted material has taken place.   As we will
discuss,  whether or not mixing has taken place has important consequences
on the secular evolution of the WD.   Therefore, in this review we distinguish between those systems in which the accreted 
material does not mix with core material and those that do mix with core
material.  We will also discuss the effects of the other parameters.

Recent reviews of the properties and evolution of CVs can be found in 
\citet{warner_1995_bk, knigge_2011_ab, knigge_2011_ac, knigge_2011_aa}.
Recent reviews of the Classical Nova phenomena can be found in \citet[][hereafter, G98]{gehrz_1998_ab},
 \citet{warner_1995_bk}, \citet[][and references therein]{ hernanz_2002_bk},
 \citet{hernanz_2008_aa}, and \citet[][hereafter, S08]{starrfield_2008_cn}. 
In the next section, we present the basic physics of a thermonuclear runaway and the
importance of the $\beta^+$-unstable nuclei.  We follow that with
sections on: the initial conditions,  the Single Degenerate (SD) scenario for 
the progenitors of Supernovae of Type Ia, the Classical Nova Outburst, a discussion
of future work, and end with a Summary and Conclusions.

\section{The Basic Physics of a Thermonuclear Runaway}

Hydrodynamic simulations show that the consequences of accretion
from the secondary is a growing layer of hydrogen-rich gas on the
WD.   When the deepest layers of
the accreted material have become both hot and electron
degenerate, a TNR occurs near the base of
the accreted layers.   For the physical conditions of temperature
and density that occur in this environment, nuclear processing
proceeds by hydrogen burning, first from the proton-proton chain
and, subsequently, via the carbon, nitrogen, and oxygen (CNO)
cycles. If there are heavier nuclei present in the nuclear burning
shell, then they will contribute significantly to the energy production and the resulting
nucleosynthesis.  Simulations of this evolution for the classical
nova (CN) outburst, designed to fit the observed
properties of the ONe nova V1974 Cyg found that changes in the
nuclear reaction library and opacities caused important changes in
the results \citep[][and references therein]{starrfield_2000_apjsupp}.  
More recently, \citet[][hereafter, S09]{starrfieldpep09} redid some of those calculations with the
Iliadis (2005, priv. comm.) reaction library and the \citet{hix_1999_ab} network solver and found that all previous work had
neglected the $pep$ reaction ($p + e^{-} +p \rightarrow d + \nu$:
\citep{schatzman_1958_aa, bahcall_1969_aa}) which, when included,
changed the ejecta abundance predictions (S09).   This reaction has a 
density squared dependence and, while not important for energy
production in the Sun, in the outer layers of a WD the density 
can reach to $10^4$ gm cm$^{-3}$ which significantly increases the energy
production at a given temperature and density.  The increased
energy production causes the temperature to increase faster for
a given mass accretion rate and the TNR occurs earlier with
less mass accreted.  A smaller amount of accreted mass implies
a lower peak temperature and, thus, the nucleosynthesis predictions change
just based on the inclusion of this one nuclear reaction rate.
  
Nevertheless, while the proton-proton chain is important during the accretion
phase of the outburst (before the rise to the TNR), during which time the amount of accreted
mass is determined, it is the CNONe cycle reactions and,
ultimately, the hot CNO sequences that power the final stages and
the evolution to the peak of the TNR.   Energy production and
nucleosynthesis associated with the CNO hydrogen burning reaction
sequences impose interesting constraints on the energetics of the
runaway.  In particular, the rate of nuclear energy generation at
high temperatures (T $>$ 10$^{8}$K) is limited by the timescales
of the slower and temperature insensitive positron decays,
particularly $^{13}$N ($\tau_{1/2}$ = 600s), $^{14}$O
($\tau_{1/2}$ = 102s), and $^{15}$O ($\tau_{1/2}$ = 176s).  The
behavior of the positron decay nuclei holds important implications
not only for the nature and consequences of CN outbursts where
the accreted and core material are mixed but also for the simulations
where there is no mixing.  For
example, significant enhancements of envelope CNO concentrations
are required to insure high levels of energy release on a
hydrodynamic timescale (seconds for WDs) and thus produce a
violent outburst \citep[][hereafter, S98; S09]{starrfield_1989_bk, starrfield_1998_aa}.

The large abundances of the positron decay nuclei, at the peak of
the TNR, have important and exciting consequences for the further
evolution of the TNR.  (1) Since the energy production in the CNO cycle comes
from proton captures, interspersed by $\beta^{+}$-decays, the rate
at which energy is produced, at temperatures exceeding 10$^{8}$K,
depends only on the half-lives of the positron decay nuclei and
the numbers of CNO nuclei initially present in the envelope. (2)
Since near peak temperature in the TNR convection operates throughout the entire accreted envelope,
unburned CNONeMg nuclei are carried into the nuclear burning region, when the
temperature is near maximum, and the nuclear
reactions operate far from equilibrium. (3) Since the convective
turn-over time scale is $\sim 10^{2}$ sec near the peak of the
TNR, a significant fraction of the positron decay nuclei are able
to reach the surface without decaying and the rate of energy
generation at the surface can exceed 10$^{13}$ to 10$^{15}$ erg
gm$^{-1}$ s$^{-1}$ (depending upon the enrichment).  (4) These
same nuclei decay when the temperatures in the envelope have
declined to values that are too low for any further proton
captures to occur, yielding isotopic ratios in the ejected
material that are distinctly different from those ratios predicted
from studies of equilibrium CNONeMg burning. (5) Finally, the
decays of these nuclei provide an intense heat source throughout
the envelope that helps eject the material from off the WD.
Theoretical studies of this mechanism show that sufficient energy
is produced, during the evolution described above, to eject
material with expansion velocities that agree with observed values
and that the predicted bolometric light curves for the early
stages are in reasonable agreement with the observations
(S89; S98; S08).  Finally, the $\beta^{+}$-decay heating of the outermost
regions of the WD envelope reduces the temperature gradient and,
in turn, curtails convection in the surface layers. The growth of
convection from the burning region to the surface and its
subsequent retreat in mass, as the envelope relaxes from the peak
of the runaway on a  thermal timescale, implies that there should
exist variations in the elemental and isotopic
abundances in the ejected gases.

\section{Initial Conditions for the Outburst}

A short description of the history of the development of the TNR hypothesis for the
outburst is given in Starrfield (1989) and will not
be repeated here.  One of the important developments since that
review have been the calculations of the amount of hydrogen-rich
material that can be accreted before the TNR is triggered.  In the
1980's there were both analytic \citep{fujimoto_1982_ab, fujimoto_1982_aa} and
semi-analytic \citep{macdonald_1983_aa} calculations to determine the
amount of material that could be accreted onto a WD before the TNR
occurred. Since that time, there have been a number of studies of
accretion onto WDs using Lagrangian hydrostatic or hydrodynamic
computer codes which follow the evolution of the material as it
gradually accretes onto the WD. These calculations show that the
amount of material accreted onto the WD is a function of the WD
mass, luminosity, the composition of the accreted matter, and the
mass accretion rate (\.M).  If
the mixing of the accreted material with the core occurs during
the accretion process, then the core nuclei added to the accreted
layers will affect the amount of material accreted prior to the
TNR by increasing the opacity and trapping the heat from the
nuclear burning regime in those layers where the heat is produced (S98).
Although there have been a
number of multi-dimensional studies investigating mixing during the
nova outburst \citep[and references therein]{glasner_1997_aa, 
kercek_1998_aa,glasner_2007_aa, casanova_2010_aa,casanova_2011_aa, casanova_2011_ab}, 
they are limited by CPU time considerations to following the evolution only near the peak of the
outburst.  If mixing actually occurs much earlier or much later in the evolution, then we still
need one-dimensional calculations to suggest the answer. 

Hydrodynamic studies show that most of the time to the TNR is spent and most of the
mass is accreted during the phase when the principle nuclear
burning process is the proton-proton chain \citep[S89; S98;][S08]{yaron_2005_aa}.   Therefore, there is a
competition between the radiative diffusion time to the surface (convection does not start until just
before the peak of the TNR) and the energy production which depends both on the temperature ($\sim$T$^{4-6}$) and the 
hydrogen mass fraction ($\epsilon \sim X^{2}$).  In addition, because these layers are degenerate, electron
conductivity can transport some of the energy into the interior.  Nevertheless, it was the original hydrodynamic
studies that showed as long as the radiative opacity was
``small'',  most of the energy produced in the nuclear burning regime 
at the bottom of the accreted layers was
transported to the surface and radiated.  Therefore,  the
temperature in the nuclear burning region increased
slowly during the proton-proton phase of accretion.   
Thus, the hydrodynamic studies showed that the
amount of mass accreted during the proton-proton phase depended on
the opacity (metallicity) of the material (S09).

If we assume that no mixing has occurred between the core and
accreted layers during the proton-proton phase, 
then increasing the metallicity of the infalling matter results in an
increase in its opacity. The increased
opacity traps more of the heat (produced by compression and
nuclear burning in the deeper layers of the accreted material)
in the region where it is produced and thus the
temperature increases faster per unit accreted mass then for
material with a lower opacity (metallicity).  In contrast,
lowering the metallicity by accreting material representative of
the LMC (one-third Solar metallicity or less), reduces the
opacity and increases the rate of heat transport out of the nuclear
burning layers.  As a result, the temperature increases more slowly and
more material is accreted.  A more massive accreted layer
causes a higher density at the bottom and the explosion becomes
more violent \citep{starrfield_1999_lmcaa, jose_2007_aa}. This result is in agreement
with the observations of novae in the LMC \citep{dellavalle_1992_ab, dellavalle_1994_aa}.

If, however,  the accreting material mixes with core material
during this phase, either by shear mixing \citep{sparks_1987_aa, kutter_1987_aa, rosner_2001_aa, alexakis_2004_ab}
or by elemental diffusion \citep{prialnik_1984_aa, kovetz_1985_aa}, the addition of heavy nuclei to the accreted
layers will increase their opacity.  The opacity increase
will reduce the amount of material accreted before
the onset of the TNR and, thereby, the amount of material ejected
during the outburst.  Given that the theoretical predictions of the
amount of material ejected during the outburst are already far lower then
the observations (S08), increasing the metals in the accreted layers by
early mixing exacerbates this disagreement.

There is an interesting corollary to this point.  The OPAL
opacities  \citep{rogers_1994_aa} and \citet{iglesias_1996_aa} were improved over previous
opacities by increasing  the number of atomic
energy levels and improving the treatment of line
broadening. This had the effect of increasing the opacities for the same densities and
temperatures even without changing the abundances.   In order to study this effect, we 
included the latest OPAL opacities \citep{rogers_1992_aa, rogers_1994_aa, iglesias_1996_aa} 
in NOVA (S98 and references therein) and 
calculated new evolutionary sequences
in order to simulate the outburst of V1974
Cyg (Nova Cyg 1992).  The increased opacities had
profound effects on the simulations.  Because the new opacities
were larger than those we had been using (the \citet{iben_1975_aa} fit to
the \citet{cox_1970_aa,cox_1970_ab} opacities), we
found that the heat from the nuclear reactions was trapped more
effectively in the layers where it was produced.  Our simulations ejected a factor of ten less mass than was inferred
from observations (S98). This discrepancy was also found in a study of
accretion onto CO and ONe WDs {\citep{jose_1998_aa,jose_1999_aa}.
One possible solution to this problem involves mixing of the accreted hydrogen-rich material into a
residual helium-rich shell which is the remnant of previous outbursts and subsequent quiescent hydrogen burning
\citep{krautter_1996_aa}.  The increased mass fraction of helium will reduce the opacity and 
allow a larger fraction of the energy produced in these layers to be transported to the
surface and radiated away.   Simulations with increased helium are in progress (Starrfield et al. 2012, in prep.)

\citet{prialnik_1982_aa} were the first to show that there was a
strong effect of the rate of mass accretion on the ignition mass.
They found that increasing the rate of mass accretion increased
compressional heating and, thereby, caused the temperatures in the
accreted layers to rise more rapidly (per given amount of accreted
mass) than for lower mass accretion rates.  The observed mass
accretion rates of $\sim10^{-9}$ M$_\odot$ yr$^{-1}$  \citep{townsley_2002_aa, townsley_2004_aa}
resulted in much smaller amounts of
material being accreted when compared to simulations where the
rate of mass accretion was a factor of 10 to 100 times smaller.
While it has also been suggested that increasing the mass accretion rate
much above $10^{-8}$M$_\odot$ yr$^{-1}$, on low luminosity WDs,
would cause extremely weak flashes, new simulations of accretion (with
no mixing of accreted with core material) 
find that even mass accretion rates as high as  $2 \times 10^{-6}$M$_\odot$ yr$^{-1}$ 
still result in TNRs and the expansion of the outermost layers to $10^{12}$cm or larger \citep{starrfieldBA12}.
\citet{starrfieldBA12} also report that TNRs occurred on WDs with
masses as low as $0.4$M$_\odot$.

Next we turn to the idea that there exists 
a ``steady burning'' regime where the infalling material burns at exactly
the same rate as it is accreted \citep{paczynski_1978_aa, Sion_1979_aa,
Iben_1982_ab, fujimoto_1982_ab,fujimoto_1982_aa, yoon_2004_aa}.  
The steady burning assumption is that only a WD accreting at a specific \.M can grow in mass.
Otherwise, it either suffers nova explosions (lower \.M) or rapidly expands, fills its Roche Lobe, and shuts off accretion (higher \.M).  
The assumption of steady burning is
relevant to the the existence of the Supersoft X-ray Binary Sources (SSS) discovered by
ROSAT \citep[][]{Trumper_1991_aa}.  The SSS sources have
luminosities of L$_* \sim 10^{37-38}$erg sec$^{-1}$
and effective temperatures ranging from $3 - 7
\times 10^5$K.  
\citet[hereafter, V92]{vandenheuvel_1992_aa} first suggested that they could be SN Ia
progenitors because
their luminosities implied that they were in the steady burning 
regime and that it was possible that the mass of the WD could be growing toward the Chandrasekhar
limit  \citep[see also][]{branch_1995_aa, Kahabka_1997_aa}.   Nevertheless, except for
the pioneering work of \citet{Sion_1979_aa} (see also
\citet{Iben_1982_ab,cassisi_1998_aa, yoon_2004_aa}), no stellar
evolution calculations had been done for massive WDs (M$_*  > 1.3$M$_\odot$) accreting at
high mass accretion rates for the sufficiently long times required to
test the steady burning hypothesis.  Calculations had been done for lower
mass CO WDs (M$_*  \sim 0.8$M$_\odot$) by \citet{Iben_1982_ab} and \citet{Sion_1994_aa}, but the
luminosities and effective temperatures of their simulations were too
low to agree with the observations of SSS such as CAL 83 or CAL 87 and
it was not clear that the WDs that they studied would reach the
Chandrasekhar Limit.  In a hydrodynamic study using NOVA
\citep[hereafter S04]{starrfield_2004_aa}, we reported that accretion
of Solar material onto hot ($2.3 \times 10^{5}$K), luminous (30L$_\odot$), 1.25M$_\odot$
and 1.35M$_\odot$ CO WDs, using a large range in mass accretion
rates, caused the accreted material to burn quiescently in the surface
layers and the WD grew in mass toward the Chandrasekhar Limit.  This
work was criticized by \citet{nomoto_2007_aa} and \citet{shen_2007_aa}
because they could not reproduce our results using {\it static}
stellar models.  In addition, \citet{nomoto_2007_aa} claimed that
the surface mass zones were too large.  Calculations of solar mass
accretion onto low luminosity WDs has now been done by \citet{starrfieldBA12} and
they find that steady burning as described above does not exist 
\citep[see also][]{idan_2012_aa}.  We discuss
this result later in this review.

Another important parameter, for a given WD mass and mass
accretion rate, is the luminosity of the underlying WD.  It has
been found that as the luminosity of the WD declines, the amount
of accreted material increases (S98).  This is because the energy
radiated by the underlying WD also
heats the accreted layers.  In fact, it is this heating plus compressional
heating as accretion progresses that causes the deepest accreted
layers to finally reach nuclear burning temperatures. 
As the luminosity decreases, this heat source becomes
less important, the accreting layers stay cool for a longer
period of time, and more mass is accreted.

The final parameter that affects the amount of material that is
accreted prior to the TNR is the mass of the WD.  All other
parameters held constant, the accreted mass is inversely proportional to the mass of
the WD \citep[][and references therein]{macdonald_1985_aa} and numbers showing this are
given in S89.  Equation 1 illustrates how the ignition mass can be estimated.

\begin{equation} P_{\rm crit} = {{\rm G M}_{\rm WD} {\rm
M}_{\rm ig} \over 4 \pi {{\rm R}_{\rm WD}^4}}
\end{equation}

P$_{\rm crit}$ is assumed to be $\sim 10^{20}$ dyne cm$^{-2}$ and
a mass-radius relation for WDs gives the ignition mass, M$_{\rm
ig}$.  Equation (1) is obtained by realizing that a critical
pressure must be achieved at the bottom of the accreted layers
before a TNR can occur \citep[][G98]{fujimoto_1982_ab,fujimoto_1982_aa}.  Note,
however, that the actual value of the critical pressure is also a
function of WD composition and rate of accretion (S89).  
If one assumes the above value for the pressure, then the
amount of accreted mass can range from less than
$10^{-5}$M$_\odot$ for WDs near the Chandrasekhar Limit to values
exceeding $10^{-3}$M$_\odot$ for 0.4 M$_\odot$ WDs. In addition,
because the surface gravity of a low mass WD is smaller than that
of a massive WD, the bottom of the accreted layers is considerably
less degenerate at the time the TNR occurs. Therefore, the peak
temperature, for a TNR on a low mass WD, may not even reach
$10^8$K so that no interesting nucleosynthesis occurs.

\section{The Single Degenerate Scenario for Supernova Ia progenitors}

The relationship between accretion onto WDs, via accretion from a non-degenerate secondary, and Supernova Ia (SN Ia) explosions is
designated the single degenerate scenario (SD).  It is one of the two major suggestions 
for the objects that explode as a SN Ia, the other being the double degenerate (DD) scenario.  In 
the standard paradigm SD scenario, a WD in a close binary system accretes material from its companion and grows to the 
Chandrasekhar Limit.  As it nears the Limit, it first convectively ``simmers'' in the core and then an explosion occurs.   
In contrast, the double degenerate scenario (DD) requires the merger or collision of two WDs to produce the observed explosion.   
While for many years the SD scenario was the more prominent, a number of concerns led to major
efforts to better understand the DD scenario.  However, the SD scenario is
capable of explaining most of the observed properties of the SN Ia explosion via the delayed detonation
model  \citep[and references therein]{khokhlov_1991_aa, kasen_2009_aa,  woosley_kasen_11_a, howell_2009_ab}.  Reviews 
of the various proposals for SN Ia progenitors \citep{branch_1995_aa},  
producing a SN Ia, and the implications of their explosions can be found in
 \citet{hillebrandt_2000_aa},
\citet{leibundgut_2000_aa,leibundgut_2001_aa},  
\citet{nomoto_2003_aa}, and \citet{howell_2011_aa}. 
 
Recently, the well studied outburst of SN 2011fe in M101 showed that the star that exploded was likely a carbon-oxygen (CO) 
WD \citep{pnugent11} with a companion that was probably a main sequence star 
\citep{weidongli_2011_aa,bloom_2012_aa} although EVLA \citep{chomiuk_2012_aa} and optical \citep{bloom_2012_aa}
observations have ruled out most types of CVs.  In addition, \citet{schaeferpag_2012_aa} find no
star at the ``center'' of a SN Ia remnant in the LMC but \citet{edwardspag_2012_aa} find a large
number of stars near the ``center'' of a second LMC SN Ia remnant and they state that they cannot rule out a CV as the
progenitor.  Nevertheless, \citet{schaeferpag_2012_aa} claim that they rule out the
SD scenario or, as is more likely, the secondary in
their LMC remnant was fainter than their detection limit. 
Even more recently, \citet{dilday_2012_aa} claim that PTF 11kx was a Symbiotic system that
exploded as a SN Ia implying strongly that there are multiple SN Ia channels.   Further support for the SD scenario, comes from 
observations of V445 Pup (Nova Pup 2000) which imply that it was a helium nova (helium accretion onto a WD) because there
were no signs of hydrogen in the spectrum at any time during the outburst but there were 
strong lines of carbon, helium, and other elements \citep[and references therein]{woudt_2005_aa, woudt_2009_aa}.   Because it was extremely luminous before the outburst, the secondary is thought to be a hydrogen deficient carbon star \citep[and references therein]{woudt_2009_aa}.   
Since one of the defining characteristics of a SN Ia explosion
is the absence of hydrogen or helium in the spectrum at any time during the outburst or decline,  the existence of V445 Pup
implies that mass transferring binaries exist in which hydrogen is absent at the time of the explosion. 

Therefore, in order to study the SD scenario, we used our one-dimensional (1-D), implicit, hydrodynamic, 
computer code (NOVA) to study the accretion of Solar composition material (Lodders 2003) onto WD masses of 0.4M$_\odot$, 0.7M$_\odot$, 1.0M$_\odot$, 1.25M$_\odot$, and 1.35M$_\odot$.  
We used two initial WD luminosities  ($4 \times 10^{-3}$ L$_\odot$ and $10^{-2}$L$_\odot$) and seven mass accretion rates ranging 
from $2 \times 10^{-11}$M$_\odot$ yr$^{-1}$ to $2 \times 10^{-6}$M$_\odot$ yr$^{-1}$.   
We  used an updated version of NOVA (S09, and references therein) that includes a nuclear
reaction network that has now been extended to 187 nuclei (up to $^{64}$Ge).  
 NOVA also uses the OPAL opacities
 \citep[][and references therein]{iglesias_1996_aa},
 the Iliadis (2005, priv. comm.) nuclear reaction rates,
recent equations of state \citep{timmes_1999_aa, timmes_2000_ab}, and new nuclear reaction network solvers \citep{hix_1999_aa, timmes_1999_ab}. NOVA can now follow CN explosions past the first outburst on the WD \citep{starrfield_2004_aa}
and uses more mass zones in the calculations.  It also includes the new \citet{arnett_2010_aa} algorithm for mixing-length convection
and the Potekhin electron conductivities described in \citet[][and references therein]{cassisi_2007_aa}.  
These improvements have had the effect of changing
the initial structures of the WDs so that they have smaller radii and larger
surface gravities which produce quantitative but not qualitative changes in our CN simulations (Starrfield et al. 2012, in prep.).

\begin{figure}[htb!]
\centering
\includegraphics[scale=0.4]{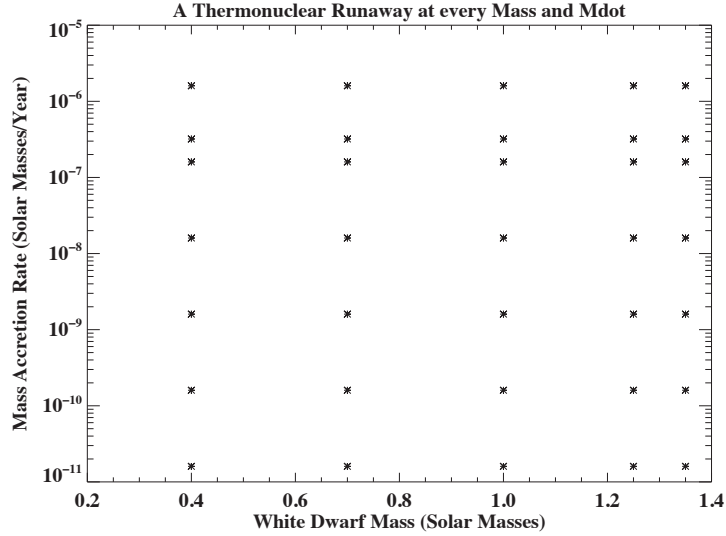}
\caption {\small WD mass vs. Log \.M for each of the evolutionary sequences that we calculated.  
Since each point represents the two initial luminosities we used, there are
70 sequences shown here.  Each of these
exhibited a TNR.  In no case did ``steady burning'' occur.}
\end{figure}

\begin{figure}[htb!]
\centering
\includegraphics[scale=0.4]{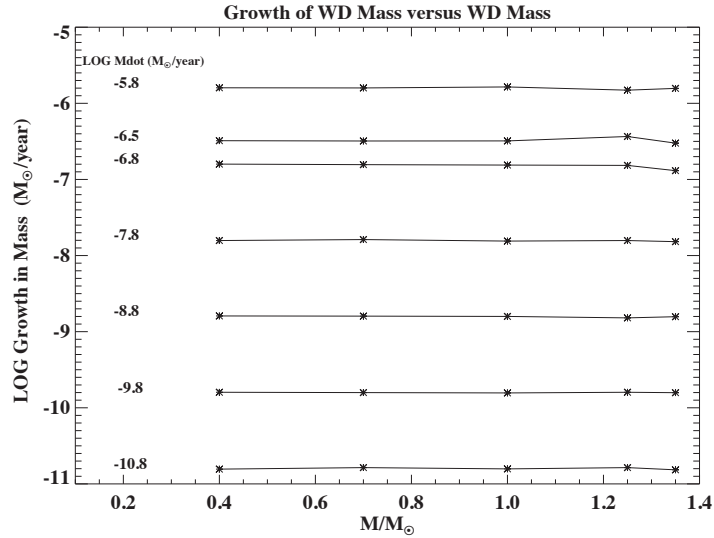}
\caption {\small The log of the difference between the mass accreted and the mass lost.  
We display the log of the growth in mass  (in units of M$_\odot$ yr$^{-1}$)
 as a function of WD mass for each of our simulations.  Each point is the amount of accreted (less ejected) mass divided by the time to 
 reach the TNR for the given simulation.  The lines connect the points for the same \.M and we give the log of  \.M along a column on
 the left of the figure. }
\end{figure}

\begin{figure}[htb!]
\centering
\includegraphics[scale=0.4]{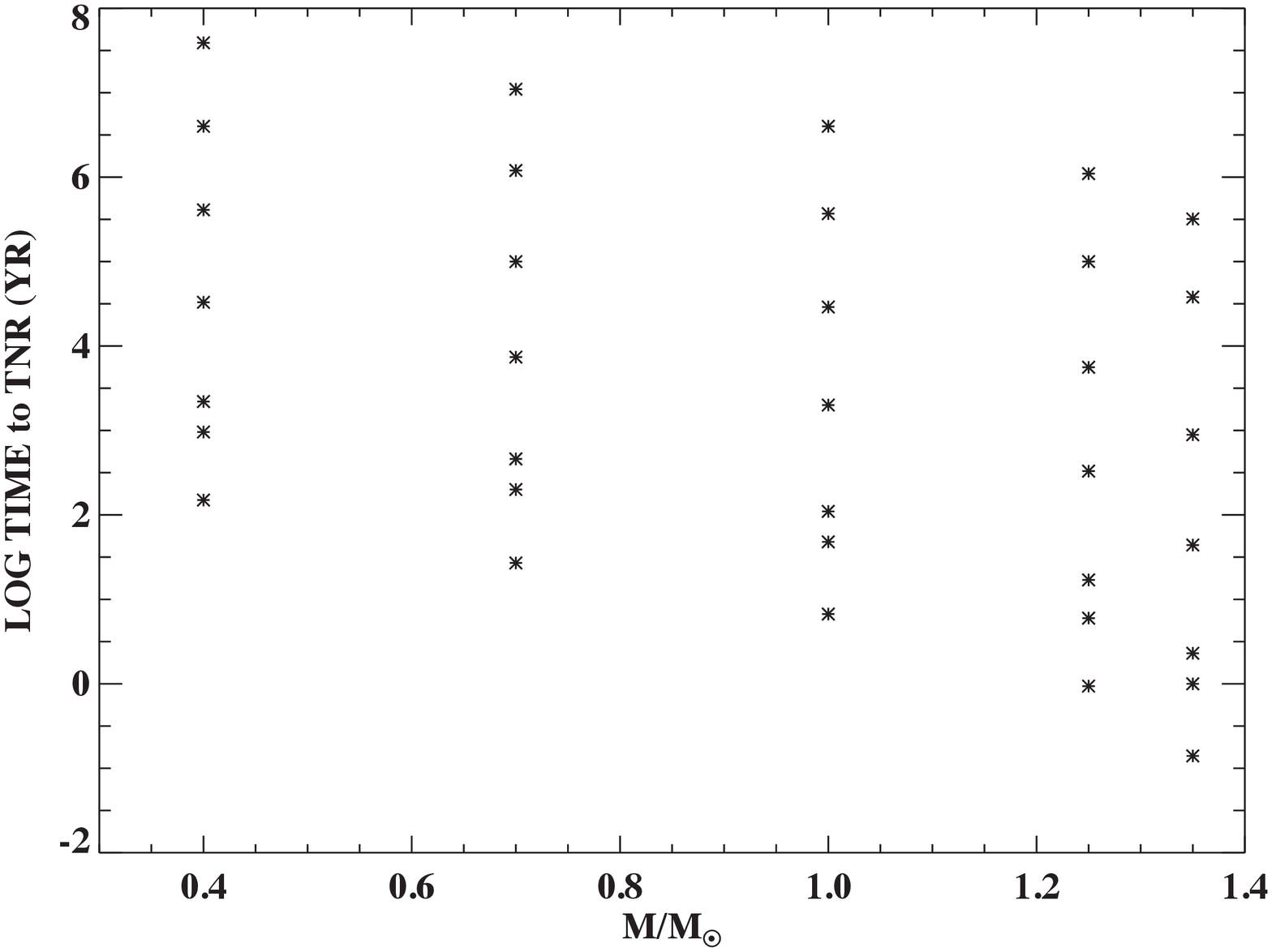}
\caption {\small The log of the accretion time to the TNR as a function of WD mass.  Each of the data
points is for a different \.M and the value of \.M increases downward for each WD mass.  The value
of \.M is given in Figure 1.  The accretion time,
 for a given \.M decreases with WD mass because it takes less mass to initiate the TNR as the WD mass
 increases.}
\end{figure}

Fig. 1 shows the results for all 70 simulations that we have done (each data point represents the two initial
luminosities).  In all cases we obtain a TNR which, for some simulations, ejects some material, and after some
evolutionary time may cause the WD radius to grow to $\sim 10^{12}$cm.  
These
fully implicit, time-dependent, calculations show that the sequences exhibit the \citet{schwarzschild_1965_aa} thin shell instability
which implies that steady burning does not occur.  An expanded study of the stability of thin shells can be found in \citet[][]{yoon_2004_aa}
who investigated
(among other studies) the accretion of hydrogen-rich material onto a WD.  Using their results, we find that our sequences begin in a stable region 
(see their Fig.  8 and Fig. 11) but with continued accretion evolve into instability.   
We also find that low mass WDs do not eject any mass while the high mass WDs do eject 
a small fraction of the accreted material (a maximum of $\sim4\%$ for the 1.25M$_\odot$  sequences but ranging down to $\sim0.1\%$ for the 0.7M$_\odot$ sequences).  Therefore, the WDs are growing in mass as a result of the
accretion of Solar material and assuming no mixing with core material.  (This is not the case for CNe which show sufficient core and accreted
material in their ejecta that the WD must be losing mass as a result of the outburst.)  
We identify these accreting systems with those CVs (Dwarf, Recurrent, Symbiotic
novae) that show no core material either on the surface of the WD or
in their ejecta.  Our results could explain the findings of
\citet{Zorotovic_2011_aa} who report that the WDs in CVs are growing in
mass.  In addition, the best studied Dwarf Novae have WD masses larger
than the canonical value of $\sim$0.6M$_\odot$.  These are 
U Gem \citep[1.2M$_\odot$:][]{Echevarria_2007_aa}, 
SS Cyg \citep[0.8M$_\odot$:][]{Sion_2010_SSCyg_aa}, 
IP Peg \citep[1.16M$_\odot$:][]{copperwheat_2010_aa}, and 
Z Cam \citep[0.99M$_\odot$:][]{Shafter_1983_aa}.  Therefore, it is possible
that some Dwarf Novae could be SN Ia progenitors.

We also show the accretion time to TNR for all our sequences (Fig. 3).  Clearly,
as the WD mass increases, the accretion time decreases for the same \.M. This is because
higher mass WDs initiate the TNR with a smaller amount of accreted mass.  Given the
existence of Recurrent Novae and Symbiotic Novae with recurrence times ranging from a few years (U Sco)
to about 20 years (RS Oph) or longer (T Pyx, V407 Cyg, and T CrB),  Fig. 3 shows that it is possible for
Recurrent Novae to occur on WDs with masses as
low as 0.7M$_\odot$.  Therefore, although it is often claimed that only the most massive 
WDs have recurrence times short enough to agree with the observations of 
Recurrent Novae, this plot shows that this is not the case and basing WD mass determinations
of Recurrent Novae on short recurrence times is incorrect.  We also
note that it is possible for a Recurrent Nova outburst to occur on a high mass WD for an extremely broad range of \.M. 

\begin{figure}[htb!]
\centering
\includegraphics[scale=0.4]{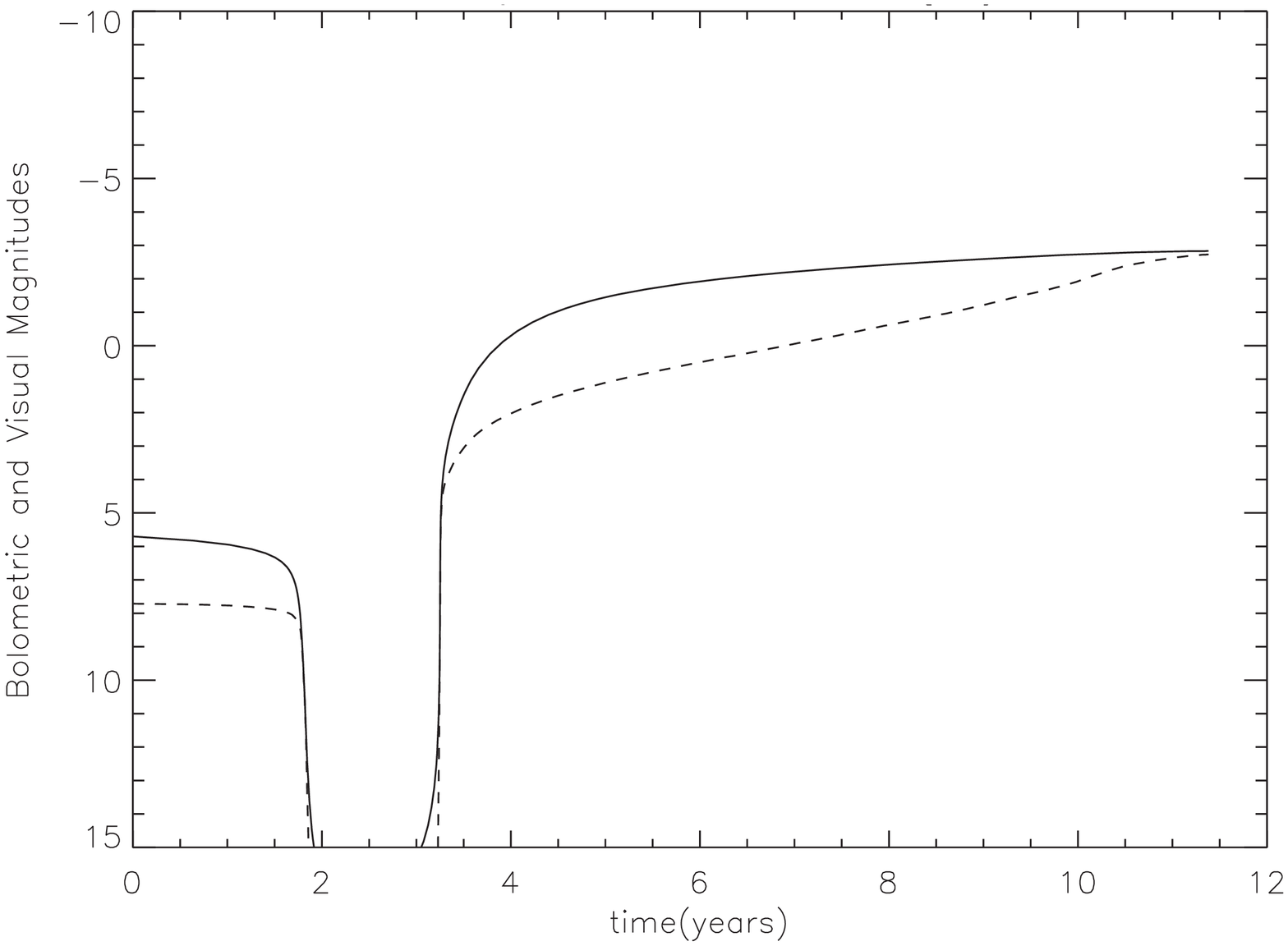}
\caption {\small The light curve for one of our sequences. The solid line is the Bolometric Magnitude and the
dashed line is the Visual Magnitude.  This sequence had a WD mass of 0.4M$_\odot$ and \.M of
$2 \times 10^{-9}$M$_\odot$ yr$^{-1}$.  Note that it takes nearly 5 years to reach maximum light
in the visual  The calculation is stopped when the outer radii reach $10^{12}$cm but no material has been
ejected. }
\end{figure}

\begin{figure}[htb!]
\centering
\includegraphics[scale=0.4]{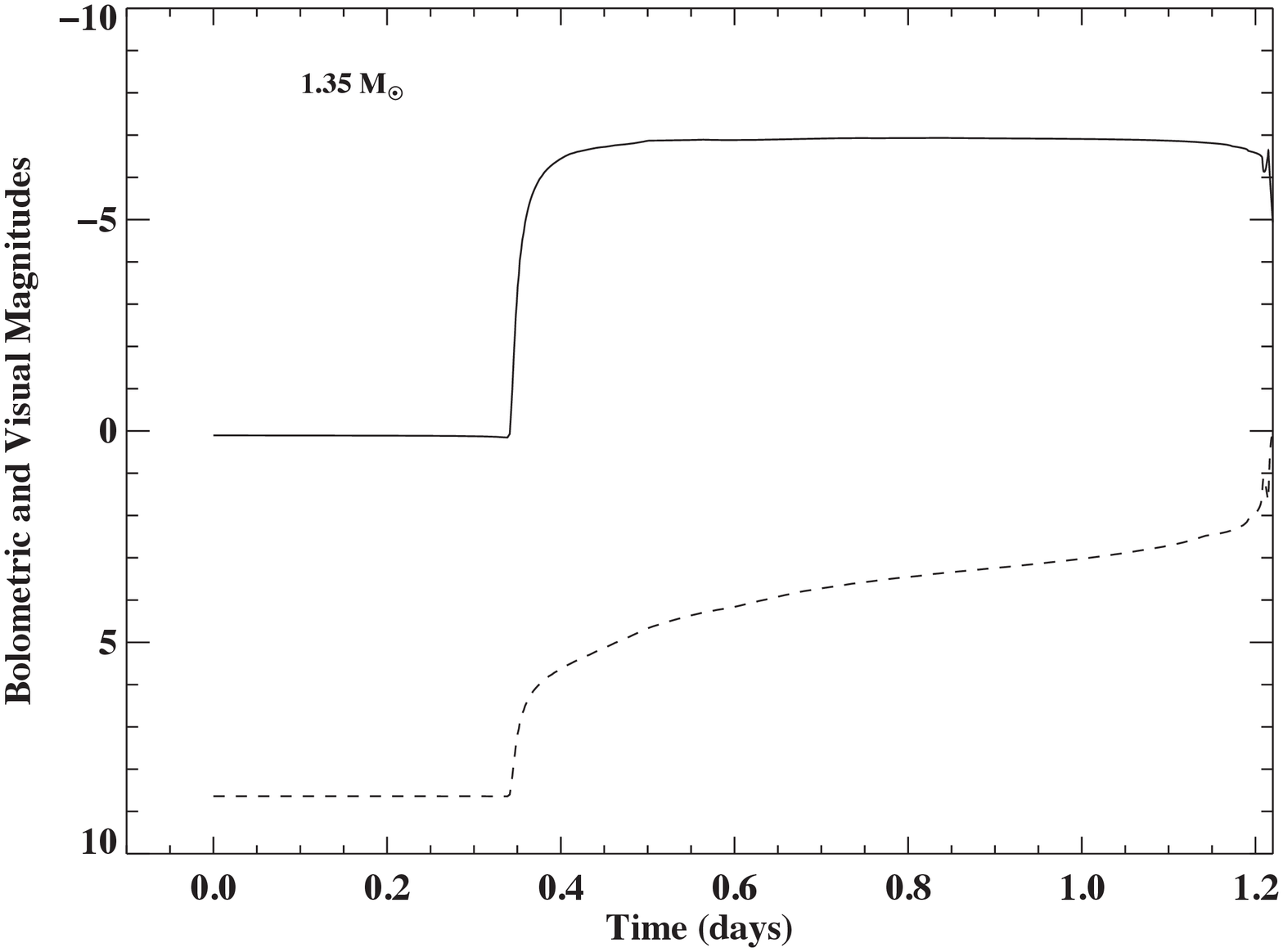}
\caption {\small This figure shows the light curve for a sequence with a mass of 1.35M$_\odot$ and an 
\.M of  $3 \times 10^{-7}$M$_\odot$ yr$^{-1}$.  We chose this rate to show that
choosing an \.M in the middle of the ``steady burning'' regime still results in a TNR.   
Note that on this WD it takes less than one day to reach maximum. A small amount of material was
ejected.}
\end{figure}

Finally, in Fig. 4 and Fig. 5 we show the light curves for two of our simulations.  The WD mass is 0.4 M$_\odot$ in Fig. 4
and 1.35 M$_\odot$ in Fig. 5.  The mass accretion rate we used for Fig 4. is $2 \times 10^{-9}$M$_\odot$yr$^{-1}$ and for Fig. 5 we used 
$3 \times 10^{-7}$M$_\odot$yr$^{-1}$.  We chose the latter value because it is in the center
of the steady burning mass accretion regime.  The evolution of the bolometric
magnitude is the solid line in each plot and the V magnitude is shown as the dashed line.   We stop the plot when
the outer radius reaches $\sim 10^{12}$cm and the expanding material has become optically thin.  The lower
mass WD takes years to evolve while the higher mass WD takes only days.  
We find that as the mass of the WD increases, the time
scale to the TNR decreases.  The initial decline before the final rise in Fig. 4 is caused by the conversion of some of the
internal energy, produced by ongoing nuclear burning near the surface, being transformed into the 
potential energy necessary for the material to climb out of the 
deep gravitational well of the WD.  The most extreme result is for the 0.4
M$_\odot$ WD for which
it takes more than one year for the expanding material to recover and begin to become more 
luminous and hotter.  Such an effect, combined with the interaction with the accretion disk 
(and possibly the secondary) might be responsible for some of the pre-maximum halts seen in some
classical novae \citep{hounsell_2010_aa}.

\section{The Classical Nova Outburst}

In previous sections we discussed the results for accretion onto WDs in
which no mixing with WD core material had taken place.  If such mixing
occurs, and sufficient core material is mixed up into the accreted layers,
then a CN outburst occurs and typically ejects material
at high speeds. In fact, the importance of continued studies, both theoretical and
observational, is that observations of the CN outburst show that a nova
ejects metal enriched gas and grains during its outburst and this
material is a source of heavy elements for the Interstellar Medium
(ISM). The observed enrichment demands that mixing of the
accreted material with core material must have taken place at some
time during the evolution to the outburst.  The ejection velocities
measured for CN ejecta can exceed, in many cases, $10^3$ km s$^{-1}$
so that this material is rapidly mixed into the diffuse gas where
it is then incorporated into molecular clouds before being formed
into young stars and planetary systems during star formation.
Therefore, CNe must be included in studies of Galactic
chemical evolution as they are predicted to be the major source
of $^{15}$N and $^{17}$O in the Galaxy and contribute to the
abundances of other isotopes in this atomic mass range.

Infrared observations have confirmed the formation of carbon, SiC,
hydrocarbons, and oxygen-rich silicate grains in nova ejecta,
suggesting that some fraction of  the pre-solar grains identified
in meteoritic material \citep{zinner_1998_aa, amari_2001_aa, amari_2001_ab, jose_2004_aa,gehrz_2008_cn, pepin_2011_aa}  may come
from CNe.  The mean mass
ejected during a CN outburst is $\sim 2\times 10^{-4}$
M$_\odot$ (G98)  Using the observed nova rate of 35$\pm$11 per
year in our Galaxy \citep{shafter_1997_aa}, it follows that CNe
introduce $\sim 7\times 10^{-3}$ M$_\odot$ yr$^{-1}$ of processed
matter into the ISM. There is probably more material ejected 
than presently believed, however, and this value is a lower
limit \citep[][G98]{saizar_1994_aa}.

CNe are frequent and varied and observations of their outbursts 
provide an extensive dataset to test theories
of their explosions and their evolution.  For example, the composition
of matter ejected in a CN outburst must depend on both the amount and
composition of the material mixed up from the underlying CO or ONe
WD core plus the phase of the TNR at which this mixing occurs.
Moreover, the envelope composition, the amount of mixing from
the nuclear burning region to the surface, and the amount of material
ejected into the ISM affects the contribution
of CNe to Galactic chemical evolution.  In addition, since
core material must be mixed with accreted material during the evolution
to explosion and then ejected into space after being processed
through hot, hydrogen burning, we have a nearly unique opportunity
to determine the core abundances of WDs of various masses and
evolutionary histories.   Finally, we also note that CNe and Recurrent Nova outbursts
are the only stellar explosions for which the nuclear
physics input at present is mainly based on the direct results of laboratory experiments and no extrapolation
to lower energies is necessary. 

The hydrodynamic studies show that at least three of the
observational behaviors of the CN outburst are
strongly dependent upon the complicated interplay of nuclear
physics and convection that occurs during the final minutes of the
TNR.  These are: (1) the early evolution of the observed light
curves on which their use as ``standard
candles'' is based. (2) The observed peak luminosity of fast CNe
which is typically super-Eddington.  In some cases for as long as
two weeks \citep{schwarz_2001_aa}. (3) The composition of matter
ejected by a CN which depends on the amount and
composition of the material dredged up from the underlying CO or
ONe WD core.  Moreover, the amount of core nuclei in the ejecta implies that
the WD in a CN system is losing mass as a result of
continued outbursts and, therefore, a CN cannot be a SN Ia progenitor
\citep{macdonald_1983_aa, starrfield_2000_apjsupp}.
Theoretical studies of the CN outburst have been recently reviewed by
S08 and  \citet{hernanz_2008_aa} and here we only briefly review the
latest work with the $pep$ reaction included (S09).  

In S09 we evolved a series of evolutionary sequences on both
1.25M$_\odot$ and 1.35M$_\odot$ WDs and compared those results with
evolutionary sequences done with the $pep$ reaction not included. 
We used four different nuclear reaction rate libraries in order to determine the
effects on CN simulations of nearly 15 years of improvements in the
nuclear reaction rates.  The
most up-to-date library at that time was that of Iliadis (2005, priv. comm.).   
Fig. 6 is taken from S09 and shows the variation of temperature with time for the zone
where peak conditions in the TNR occurred in the 1.35M$_\odot$
evolutionary sequences. Here we
plot four simulations done with the $pep$ reaction
included and the \citet{anders_1989_aa} Solar abundances mixed with a 
half Solar and half ONe mixture (see S09).  The initial conditions for
all 4 sequences are the same and listed in S09.  The nuclear reaction rate library
used for each sequence is identified on the plot and in
the caption.  The reference to the library used for that
calculation is given in the caption.   The time
coordinate is arbitrary and chosen to clearly show each
evolutionary sequence.  There are clearly differences between the four
simulations.   Peak temperature drops from $\sim4.1 \times 10^{8}$ K to
$\sim3.9 \times 10^8$ K and peak nuclear energy generation
drops by about a factor of 2 from the oldest library to the newest
library ($8.4 \times 10^{17}$erg gm$^{-1}$s$^{-1}$ to $4.4 \times
10^{17}$erg gm$^{-1}$s$^{-1}$).  The temperature declines more
rapidly for the sequence computed with the oldest reaction library
\citep{politano_1995_aa} because it exhibited a larger release of nuclear energy
throughout the evolution. This causes the overlying zones to
expand more rapidly and the nuclear burning region to cool more
rapidly. In contrast, the newest library, with the smallest
expansion velocities, cools more slowly.  As expected for the increased
WD mass and gravity, the simulations at 1.35M$_\odot$ evolve
more rapidly near the peak in the TNR than those at 1.25M$_\odot$.

\begin{figure}[htb!]
\includegraphics[scale=0.4]{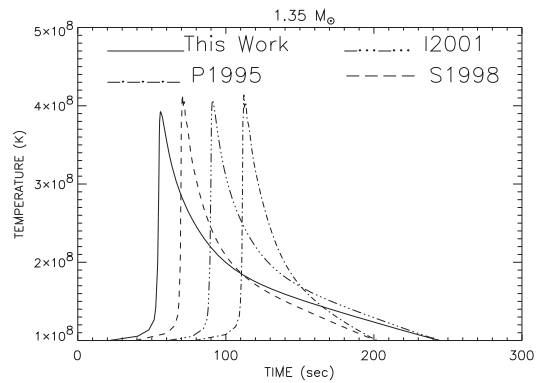}
\centering 
\caption{\small The variation with time of the temperature in
the zone in which the TNR occurs around the time of peak
temperature This zone is usually one zone above the core-envelope
interface. We have plotted the results for four different
simulations on a 1.35M$_\odot$ WD. The identification with
calculations done with a specific nuclear reaction rate library is given on the plot. 
S1998 refers to \citet{starrfield_1998_aa}, 
P1995 refers to \citet{politano_1995_aa}, I2001 refers
to \citet{iliadis_2001_aa}, and This Work refers to the calculations
done with the August 2005 Iliadis nuclear reaction rate library
and reported in S09. 
(The details of the August 2005 library are given in S09.)  The curve for each sequence
has been shifted slightly in time to improve its visibility. }
\end{figure}

In order to
more clearly show which nuclei are produced by CNe explosions, in
Fig. 7  we plot the stable, ejected nuclei compared to
the Anders and Grevesse (1989) Solar abundances (Timmes et al.
1995).  The $x$-axis is the atomic mass number.
The $y$-axis is the logarithmic ratio of the ejecta abundance
divided by the Solar abundance of the same nucleus.  The most
abundant isotope of a given element is marked by an asterisk and
isotopes of the same element are connected by solid lines and
labelled by the given element. These plots are patterned after
similar plots in Timmes et al. (1995).  They show for the 1.35M$_\odot$ simulation
done with the latest reaction rate library 
that $^{15}$N, $^{17}$O, and $^{31}$P are significantly
overproduced in CN ejecta. Other nuclei are overproduced by
factors of a thousand and could be important for CN
nucleosynthesis. 

\begin{figure}[htb!]
\includegraphics[scale=0.6]{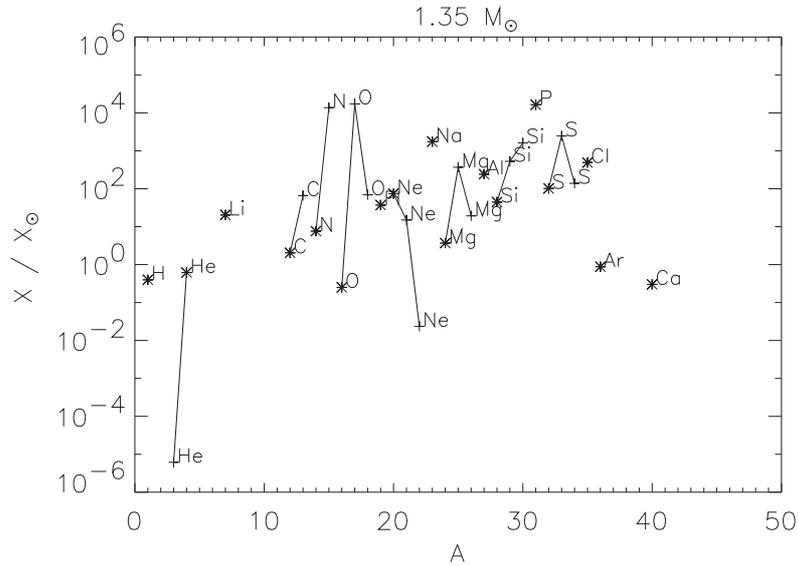}
\centering 
\caption{\small The abundances (mass fraction) of the stable
isotopes from hydrogen to calcium in the ejected material for the
1.35M$_\odot$ sequence calculated with the August 2005 reaction
rate library of Iliadis.  The $x$-axis is the atomic mass and the
$y$-axis is the logarithmic ratio of the abundance divided by the
corresponding \citet{anders_1989_aa} Solar abundance. As in
\citet{timmes_1995_aa}, the most abundant isotope of a given element
is designated by an ``$*$'' and all isotopes of a given element
are connected by solid lines.  Any isotope above 1.0 is
overproduced in the ejecta and a number of isotopes are
significantly enriched in the ejecta.}
\end{figure}

All abundances given in the following discussion are in mass fraction.
The initial abundance of $^1$H is 0.365 (half Anders and Grevesse
[1989] and half ONe and is the same abundance used in \citet[][P1995]{politano_1995_aa} and
\citet[][S1998]{starrfield_1998_aa} and its abundance declines to $\sim$0.31 in all 5
sequences. This decline of 0.05 in mass fraction results in energy production from proton captures of $\sim 2 \times
10^{46}$ erg which agrees with the values typically quoted for CN
explosions (G98; S08).  Interestingly, the abundance of $^4$He
decreases slightly as the reaction rate library is improved so
that the smallest increase in $^4$He occurs in the calculations done with
the latest library.  
These results show, as emphasized in \citet[][see also S1998 and S08]{krautter_1996_aa},
that the large amounts of
helium observed in CN ejecta (G98) implies first, that most of the
observed helium was mixed up from the outer layers of the WD by
the TNR; and second, that it was actually produced in previous CN
outbursts on the WD. This speculation is also relevant for the
large enrichments of nitrogen observed in CN ejecta (S08). The
observed nitrogen is probably $^{15}$N produced in previous
outbursts, mixed into the accreted material from the WD outer
layers, and then ejected by the CN outburst. 
Examining the behavior of the individual abundances, $^{12}$C and $^{13}$C are virtually unchanged by the updated
reaction rates. In contrast, the abundance of $^{14}$N nearly
doubles and that of $^{15}$N decreases by a factor of two going
from the first to the latest reaction rate library.  $^{16}$O also
doubles in abundance while $^{17}$O grows by a factor of 60 and
becomes the most abundant of the CNO nuclei in the ejecta.  For
this WD mass, the C/O ratio is 0.12.  The abundance of $^{18}$O
declines by nearly a factor of 5 and the abundances of $^{18}$F
and $^{19}$F also decline by large factors.

The initial abundance of $^{20}$Ne in all four sequences is 0.25
and it is depleted by a smaller amount in the
calculations done with the latest library.  The abundance of
$^{22}$Na decreases with the library update and $^{24}$Mg is
severely depleted by the TNR. In fact, all the Mg isotopes are
depleted in the calculations done with the latest library.  In
contrast, $^{26}$Al is unchanged by the changes in the reaction
rates while the abundance of $^{27}$Al drops by a factor of two.
This result implies that TNRs on more massive WDs eject about the
same fraction of $^{26}$Al as $^{27}$Al.  Contrary
to a conclusion in \citet{politano_1995_aa}, S09 finds that the amount of $^{26}$Al ejected is
virtually independent of WD mass.
All the Si isotopes ($^{28}$Si, $^{29}$Si, and $^{30}$Si) are
enriched in the calculations done with the latest library and because of the higher temperatures reached in the simulations
$^{29}$Si, and $^{30}$Si are more abundant in the 1.35M$_\odot$
simulations than in the 1.25M$_\odot$ simulations.  Other nuclei
whose abundance is largest in the calculations done with the
latest library are $^{31}$P and $^{32}$S.  These nuclei are also
more abundant at the higher WD mass because of the higher temperatures
reached in the simulations on more massive WDs.  In addition, while the
abundance of $^{33}$S is hardly dependent on the reaction rate
library, it is nearly 30 times more abundant in the calculations
done with the more massive WD.  Finally, we note that while the
ejecta abundances of $^{34}$S, $^{35}$Cl, $^{36}$Ar, and $^{40}$Ca
have all declined as the reaction rate library has been improved,
they are all produced in the nova TNR since their final abundances
exceed the initial abundances.

\section{Where do we go from here?}

The calculations that we have reported above show that we get observed behaviors both
with and without mixing of accreted material with core material.  With mixing we
get CNe and the observations imply that because of the explosion enough mass is lost from the WD that
it cannot be growing toward a SN Ia explosion.  In contrast, if no mixing occurs,
then we get TNRs that occur on timescales in agreement with Recurrent and Symbiotic
Novae but as is observed almost no mass is ejected.  Thus, we predict that these systems and Dwarf
Nova systems that contain WDs that are growing in mass could reach the Chandrasekhar
Limit and explode as a SN Ia.  The question then becomes: why do some systems mix and
some do not.   Although there have recently been a
number of multi-dimensional studies investigating mixing during the
nova outburst 
\citep[and references therein]{casanova_2010_aa,casanova_2011_aa, casanova_2011_ab}, 
as already mentioned they are limited to following the evolution only near the peak of the
outburst.  Mixing may actually occur much earlier or much later in the evolution thus further
one-dimensional calculations are necessary.  Finally, as noted earlier,  observations of V445 Pup (Nova Pup 2000)
showed no signs of
hydrogen in the spectrum at any time during the outburst 
\citep[and references therein]{woudt_2005_aa, woudt_2009_aa}
and continued studies of accretion of helium-rich material are necessary.

Because of these uncertainties and new discoveries, new simulations 
of accretion both with mixing and without mixing using
a variety of compositions are
required to better understand the secular evolution of these systems.   The simulations
done by our group will continue but with two major changes.  First, we will use MESA, as described
below, and second we will switch to a new
reaction rate library currently being constructed called STARLIB 
\citep{Lon10, Ili10a, Ili10b, Ili10c}.  In the next few paragraphs we discuss
why these changes are necessary.  

Thermonuclear reaction rates are an essential ingredient for any
stellar model. A library of experimental thermonuclear reaction rates,
based on nuclear physics input gathered from laboratory measurements,
was first published by Fowler and collaborators more than 40 years ago
\citep[and references therein]{fowler_1975_ab, caughlan_1988_aa}. 
The incorporation of Fowler's rates into stellar evolution
calculations represented a paradigm shift for the field of stellar
structure and evolution.  Subsequent work
\citep{Ang99, iliadis_2001_aa} incorporated newly measured nuclear cross
sections, but the reaction rates were still computed using techniques
developed prior to 1988.  One problem with these, and most other, published thermonuclear
reaction rates was that either a recommended value was reported without
any estimate of its uncertainty, or recommended values were published
together with ``upper limits" and ``lower limits" that had no
statistically rigorous foundation.  A solution to this problem was devised
by \citet{Lon10} and \citet{Ili10a,Ili10b,Ili10c}.  

Suppose laboratory measurements of all necessary nuclear physics
quantities have been performed.  Using the probability density
function for each nuclear physics input, one can randomly sample
over each distribution and calculate a thermonuclear reaction rate
according to the usual formalism \cite[e.g.,][]{Ili07}.  Repeating the
sampling many times provides the Monte Carlo reaction rate probability
density. Its associated cumulative distribution is then used to derive
reaction rates with a precise statistical meaning.  An example is
shown in Fig. \ref{rates} for a single resonance in a hypothetical
reaction. Randomly sampling over the measured energy and strength of
this resonance yields the reaction rate probability density
distribution, shown in red (top).  Note that the reaction rate is
well approximated by a lognormal function (black line). The lower
panel displays the corresponding cumulative distribution that can be
used to derive reaction rates as quantiles. For example, the 0.16,
0.50, and 0.84 quantiles correspond to a ``low rate", ``recommended
rate", and ``high rate", respectively, for a coverage probability of
68\%.  Other values for the quantiles can be chosen.  The important
point is that any reaction rate derived from the rate probability
density has a precisely defined coverage probability.  Therefore, for
the first time, we can provide statistically meaningful thermonuclear
reaction rates and use them to perform realistic stellar evolution and nucleosynthesis simulations.

\begin{figure} [htb!] 
\centering
\includegraphics[scale=0.4]{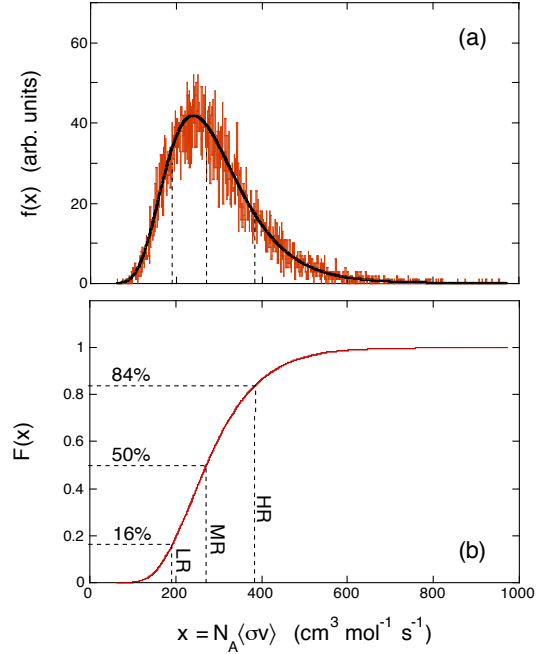} 
\caption{ \small Results of Monte Carlo calculations for a single resonance in a
hypothetical reaction at a temperature of T=0.5 GK. The resonance
parameters are $E_r=300\pm15$~keV and
$\omega\gamma=4.1\pm0.2$~eV. The reaction rate is sampled 10,000
times; (top) reaction rate probability density function, shown in
red; the black solid line represents a lognormal approximation (see
text); (bottom) cumulative reaction rate distribution; the vertical
dotted lines represent the low, median and high Monte Carlo reaction
rates, obtained from the 0.16, 0.50 and 0.84 quantiles,
respectively. }
\label{rates}
\end{figure}

Experimental Monte Carlo reaction rates are now available for 66
reactions involving stable and radioactive target nuclei in the A=14-40 range, 
including many of the key reactions for stellar
hydrogen through oxygen burning \citep{Ili10a}.  Based on these rates,
Iliadis et al.  have constructed a comprehensive, next-generation nuclear reaction
rate library for stellar modeling, called \hbox{STARLIB}. It has a
tabular format and lists, for 60 temperature grid points between
$10^6$ K and $10^9$ K,  the recommended (best) rate and an additional
parameter (the spread parameter of the lognormal rate distribution).
With the tabulated information the reaction rate probability density
function can be calculated at each temperature grid point; see \citet{Lon10}
for details.  To the experimental Monte Carlo rates Iliadis et al. have added: (i) other experimental
reaction rates from the literature for which Monte Carlo rates are not
yet available; (ii) the latest experimental and theoretical weak
interaction decay rates; (iii) the latest theoretical Hauser-Feshbach
rates calculated using the code TALYS \citep{Goriely08}; and (iv)
experimental rates for neutron-induced reactions based on the KADoNIS
v0.3 evaluation\footnote{See: \texttt{http://www.kadonis.org}.}.
For all reaction or decay rates based on theory a factor of 10
uncertainty is adopted. The STARLIB library is publicly
available and also provides the  Monte Carlo reaction rate code to the
astrophysics community \footnote{The website is
\texttt{http://starlib.physics.unc.edu/index.html}; username: 
 starlib; password: bil.rats.}.  The importance of using STARLIB in new
 simulations is that it will contain the latest rates in a form
 that is easily incorporated into stellar evolution simulations.  There are numerous rates
 that still need new measurements and our studies will help to identify those rates that
 strongly affect the nucleosynthesis.  Finally, we emphasize that using this reaction
rate library now allows us to do stellar evolution with ``error bars" in the sense that we can
easily vary the rates in STARLIB using the spread parameter, and determine the robustness
of our results.

We will combine the rates from STARLIB with MESA.  MESA\footnote{See: \texttt{http://mesa.sourceforge.net/}
and \texttt{http://mesastar.org/}.} 
\citep{paxton_2011_aa},  is a set of modules for computational
stellar astrophysics. It includes open source
libraries for a wide range of simulations in stellar astrophysics.
Among these are one-dimensional hydrodynamics, the latest opacities and equations of state, a number of nuclear
reaction rate libraries (including STARLIB), adaptive mesh refinement, element diffusion,
rotation, and pulsation.  Although it is a relatively new code, it is now being used by a large number of
researchers on a broad variety of problems.  The large 
number of people working to improve it and maintain it suggests that it is time to
move away from NOVA.

We are now using MESA + STARLIB to study the
accretion of both hydrogen- and helium-rich material onto WDs using a
broad range in both WD properties (mass, initial luminosity, mass
accretion rate, and composition) and composition of the
accreting material.  Our initial conditions have been chosen to mimic
those observed for CVs (Dwarf, Classical, Recurrent, and Symbiotic
Novae).  We have compared the initial studies done with MESA to those
already done with NOVA \citep[and references
 therein]{starrfield_2004_aa, starrfieldpep09, starrfieldBA12} using
both codes to verify and validate the results.  Using MESA produces quantitative but not qualitative
changes in our accretion simulations (Newsham, et al 2012, in prep).
In addition, while simulations with NOVA can be done with up to 400
Lagrangian zones, we are using more than 3000 zones in our studies
with MESA.  We will also include chemical diffusion and
rotation in our studies which we cannot do with the current version of
NOVA.  Including diffusion will make the results more dependent on \.M
since the lower the \.M the longer the time to
explosion and the more time for chemical diffusion to become important
\citep[and references therein]{prialnik_1984_aa, Kovetz_etal_84b,Iben_1991_aa}.

Our work in this area will, therefore, involve four major changes to our previous calculations:
(i) use MESA with STARLIB, (ii) improve the treatment of convection and include rotation and diffusion, (iii) increase the number of mass zones 
used in the simulations, and (iv) vary the composition of the accreting material.   
We are studying:
\noindent{\bf 1.} The effects of using the new Monte Carlo
reaction rates  on our simulations of Dwarf, Classical, Recurrent, and Symbiotic Novae (accretion onto
low luminosity WDs), and the Super Soft X-ray binary sources (accretion onto high luminosity WDs).
\noindent{\bf 2.} The consequences of accreting either a pure helium mixture or a hydrogen deficient helium enriched mixture onto
WDs varying the WD mass, luminosity, and prior history. 
\noindent {\bf 3.}  A new set of studies of TNRs for both CO and ONe WDs where we include the various
mechanisms that can cause mixing of accreted with core material. 
 \noindent {\bf 4.}   A Monte Carlo post-processing study of CN nucleosynthesis using temperature-density-time
trajectories from our evolutionary calculations.

\section{Conclusions}

We have reviewed our current knowledge about the
thermonuclear processing that occurs during the evolution of the
accretion onto WDs both with and without the mixing of core with accreted
material.  If the SD scenario for the progenitors of SN Ia is valid, then we require 
the growth of a CO WD to the Chandrasekhar Limit.  This, in turn, requires that
more material remain on a WD after a TNR than is ejected by the TNR.  Hydrodynamic
simulations and observations of the CN outburst, where mixing must have occurred,
show that more mass is lost than accreted by the WD and a CN cannot be a SN Ia
progenitor.  In contrast, our hydrodynamic
simulations of accretion of solar material onto WDs {\it without mixing} always produce a TNR
and ``steady burning'' does not occur.  We have studied a broad range in WD mass (0.4 M$_\odot$  to 1.35 M$_\odot$)
and find that the maximum ejected material ($\sim4\%$) occurs 
for the 1.25M$_\odot$  sequences and then decreases to $\sim 0.1\%$ for the 0.7M$_\odot$ sequences.  
Therefore, the WDs are growing in mass as a consequence of the accretion of solar material,
and as long as there is no mixing of accreted material with core material.  Finally, 
the time to runaway is sufficiently short for accretion onto most of the WD masses that we
studied that Recurrent Novae could occur on a much broader range of WD mass than heretofore
believed.

In contrast, a TNR in the accreted hydrogen-rich
layers on the {\it low} luminosity WDs in CV
binary systems, where mixing of core matter with accreted material occurs,
is the outburst mechanism for Classical,
Recurrent, and Symbiotic novae.   The differences in characteristics of
these systems is likely the WD mass and mass accretion rate.  
The importance of studying the large numbers of CNe that are 
discovered each year is that the interaction between the
hydrodynamic evolution and nuclear physics lies at the basis of
our understanding of how the TNR is initiated, evolves, and grows
to the peak of the explosion.  The high levels of enrichment of
CN ejecta in elements ranging from carbon to sulfur
confirm that there is dredge-up of matter from the core of the WD
and enable them to contribute to the chemical
enrichment of the interstellar medium.  Therefore, studies of
CNe can lead to an improved understanding of Galactic
nucleosynthesis, some sources of pre-solar grains, and the
Extragalactic distance scale.

It is recognized that the characteristics of the CN explosion are strongly 
dependent upon a complicated interplay
between nuclear physics, the $\beta^+$ limited CNO reactions, and
convection during the final stages of the TNR. The light curves,
the peak luminosities (which can exceed the Eddington luminosity),
the levels of envelope enrichment, and the composition of
CN ejecta are all strongly dependent upon the extent
and timescale of convective mixing during the explosion. The
characteristics of the outburst depend on the white dwarf mass,
luminosity, mass accretion rate, and the chemical composition of
both the accreting material and WD material.  The evolution of the outburst also depends on
when, how, and if the accreted layers are mixed with the WD core which is still unknown.  The
importance of nuclear physics to our understanding of the progress
of the outburst can be seen when we compared a series of
evolutionary sequences in which the only change was the
nuclear reaction rate library.

Finally, we described the STARLIB reaction rate library and how we will use it
with MESA to improve our understanding of accretion of material onto WDs.

\section*{Acknowledgements}

We are grateful to a number of collaborators who over the years
have helped us to better understand the nova outburst.   We have
benefitted from discussions with A. Champagne, A. Evans, R. D.
Gehrz, P. H. Hauschildt, M. Hernanz, I. Idan, J. Jos\'e, S. Kafka, J.
Krautter, A. Mezzacappa, J.-U. Ness, G. Schwarz, G.
Shaviv, S. N. Shore, E. M. Sion, P. Szkody, J. Truran, K. Vanlandingham, R. M. Wagner, M. Wiescher,
and C. E. Woodward.  SS is grateful to S. Kafka for reading and commenting on an earlier version
of this manuscript.  We gratefully acknowledge partial support from NASA and
NSF grants to our respective institutions. CI acknowledges partial support from the U.S.
Department of Energy.


\bibliographystyle{apj}
\bibliography{references_iliadis,starrfield_master}

\label{lastpage}
\end{document}